\documentclass[conference,letterpaper,9pt]{IEEEtran}
\IEEEoverridecommandlockouts

\usepackage{cite}
\usepackage{amsmath,amssymb,amsfonts}
\usepackage{algorithmic}
\usepackage{graphicx}
\usepackage{textcomp}
\usepackage{xcolor}
\usepackage[
colorlinks=true,
urlcolor=blue,
linkcolor=green
]{hyperref}
\def\BibTeX{{\rm B\kern-.05em{\sc i\kern-.025em b}\kern-.08em
    T\kern-.1667em\lower.7ex\hbox{E}\kern-.125emX}}

\usepackage{booktabs}
\usepackage{siunitx}

\newcommand{\vect}[1]{\mathbf{#1}}

\begin{document}

\title{A first-order DirAC-based parametric Ambisonic coder for immersive communications\\
\thanks{\IEEEauthorrefmark{2} work was performed while with Fraunhofer IIS.}
}

\author{
    \IEEEauthorblockN{
        Guillaume Fuchs\IEEEauthorrefmark{1},
        Florin Ghido\IEEEauthorrefmark{1}\IEEEauthorrefmark{2},
        Dominik Weckbecker\IEEEauthorrefmark{1},
        Oliver Thiergart\IEEEauthorrefmark{1}}
    \IEEEauthorblockA{\IEEEauthorrefmark{1},\IEEEauthorrefmark{2} Fraunhofer IIS, Erlangen, Germany \\
    Email: {guillaume.fuchs, dominik.weckbecker, oliver.thiergart}@iis.fraunhofer.de}
}

\maketitle

\begin{abstract}
Directional Audio Coding (DirAC) is a proven method for parametrically representing a 3D audio scene in B-format and is capable of reproducing it on arbitrary loudspeaker layouts.
Although such a method seems well suited for low bitrate Ambisonic transmission, little work has been done on the feasibility of building a real system upon it.
In this paper, we present a DirAC-based coding for Higher-Order Ambisonics (HOA), developed as part of a standardisation effort to extend the 3GPP EVS codec to immersive communications. 
Starting from the first-order DirAC model, we show how to reduce algorithmic delay, the bitrate required for the parameters and complexity by bringing the full synthesis in the spherical harmonic domain.
The evaluation of the proposed technique for coding  3\textsuperscript{rd} order Ambisonics at bitrates from 32 to 128 kbps shows the relevance of the parametric approach compared with existing solutions.
\footnotetext{A demo is avalaible at \url{https://fhgspco.github.io/fodirac_hoa/}}
\end{abstract}

\begin{IEEEkeywords}
Ambisonics, directional audio coding (DirAC), higher-order Ambisonics (HOA), spatial audio coding.
\end{IEEEkeywords}

\section{Introduction}

Ambisonics is a scene-based spatial audio format that is becoming increasingly popular in 3D audio processing due to its great flexibility, making it particularly well suited to virtual reality (VR) and augmented reality (AR) applications.
It describes the sound field in all directions whithin the inner surface of a sphere and, unlike channel-based formats, is agnostic with respect to the reproduction system and can be easily manipulated.
However, to achieve high spatial resolution, Ambisonics requires a higher order, and transmitting the resulting number of channels, which scales quadratically with the order, is prohibitive for most applications.

Several techniques for coding jointly the Ambisonic channels have been proposed, exploiting the fact that a point source at a given position in space contains large redundancy over the Ambisonic channels.
In its simplest form, static pre-matrixing of the Ambisonic channels forms beam patterns, where the resulting spatial channels in an equivalent spatial domain are more suitable for traditional stereo or multichannel perceptual coders.
This principle is used in the Channel Mapping Family (CMF) 3 of OPUS~\cite{opus_ambix}, which uses fixed matrices in pre and post-processing, but is limited in efficiency and is restricted to target low Ambisonic order at medium bitrates.
More advanced techniques rely on statistics between Ambisonic channels to derive optimal inter-channel predictions~\cite{McGrath_2019} or to decompose them into statistically orthogonal components~\cite{peters2016scene, Zamani_2017, mahe_2019, xu2021higher}. 
For example, MPEG-H~\cite{Bleidt_2017} uses SVD decomposition to separate the directional components called predominant sounds and coded as audio objects, from a residual ambient component.
These techniques offer advantages over coding the Ambisonic channels independently using a multi-mono coder, but they require bitrates significantly higher than \SI{128}{kbps} for HOA or they should confine the transmission and reproduction to lower orders.

To further reduce the bitrate, parametric coding such as Directional Audio Coding (DirAC)~\cite{pulkki2007} is an appealing option.
DirAC is an efficient parametric approach for representing a spatial audio scene by analyzing a FOA signal.
It relies on a perceptually motivated representation of the sound field parametrized by a direction of arrival (DOA) and a diffuseness measured per frequency band. 
It is built upon the assumption that at one time instant and for one critical band, the spatial resolution of the auditory system is limited to decode one cue for direction and another for inter-aural coherence.
The sound field is then reproduced in frequency domain by cross-fading two streams: a directional non-diffuse stream and non-directional diffuse stream.
DirAC has also been extended to analyze High-Order Ambisonics (HOA) in Higher-Order Directional Audio Coding (HO-DirAC)~\cite{Politis_2015}. 
It divides the sound field into sectors, wherein DirAC parameters are estimated for each of them.
HO-DirAC has been considered for HOA coding in~\cite{Hold_2024,Hold_icassp_2024}, but is outside the scope of this work since it requires more than 4 transport channels to code and thus hardly scales to the low bitrate range considered.

Although the idea of using first-order (FO) DirAC to communicate a 3D scene seems logical and has been mentioned before, no real and practical system has been proposed yet.
In~\cite{hirvonen2009perceptual}, a coding scheme is proposed for a specific teleconference scenario, which is limited to planar audio scenes and does not generalize well to other applications and general Ambisonic coding.
In this paper, we aim to demonstrate that FO-DirAC is a valid choice for low bitrate Ambisonic coding and how it can extend a communication codec like the 3GPP EVS~\cite{3GPP445} to immersive communications.
The main contributions of the work are:
\begin{itemize}
    \item Methods for adapting FO-DirAC to HOA coding and to real-time communication constraints, i.e. low algoithmic delay, low bitrates and relatively low complexity.
    \item Justifications of design choices related to the DirAC-based techniques adopted in the recent 3GPP Immersive Voice and Audio Services (IVAS)~\cite{3GPP253} for Scene Based Audio coding.
    \item Comprehensive evaluation at different bitrates of the proposed system compared with existing Ambisonic coding solutions using the legacy EVS and OPUS~\cite{rfc6716} audio communication codecs.
\end{itemize}

\begin{figure*}[t]
    \centering
    \includegraphics[width=\linewidth]{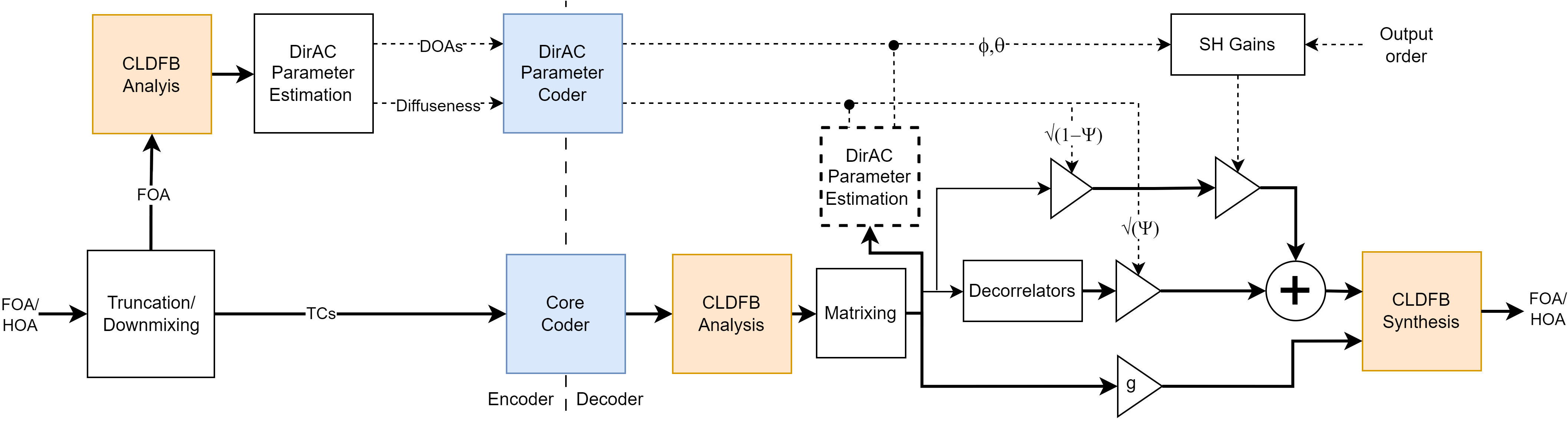}
    \caption{Block diagram of the FO-DirAC-based Ambisonic encoder and decoder.}
    \label{fig:overview}
\end{figure*}

\section{Background}

\subsection{Ambisonics format}

Ambisonics is a 3D audio format decomposing the sound field into spherical harmonics~\cite{zotter2019ambisonics}.
An audio signal arriving from a direction $(\phi,\theta)$ results in a spatial audio signal $S(\phi,\theta,t)$ which can be represented in Higher Order Ambisonics (HOA) by expanding the spherical harmonics:
\begin{equation}
    S(\phi,\theta,t)=\sum_{l=0}^{\infty} \sum_{m=-l}^{+l} Y_l^m (\phi,\theta) B_l^m (t)
\end{equation}
whereby $Y_l^m(\phi,\theta)$ being the spherical harmonics of order $l$ and degree $m$, and $B_l^m (t)$ are the Ambisonic channels.
The expansion is in practice limited up to a truncation order $l\leq H$, corresponding to $(H+1)^2$ Ambisonic channels.
With increasing truncation order $H$ the expansion results in a more precise spatial representation.
In this work, we consider orders up to $H=3$, resulting in 16 Ambisonic channels that are generally sufficient to render on standard loudspeaker layouts. 
In the following the Ambisonic Channel Number (ACN) channel order along with the normalization SN3D is adopted as in the format AmbiX~\cite{Nachbar2011AMBIXA}.

\subsection{FO-DirAC principle}


DirAC consists of two stages: analysis and synthesis.
In the analysis stage, a First-Order Ambisonic (FOA) signal (also called B-format) is transformed in a time-frequency (TF) domain.
Four components per TF-tile are obtained and can be denoted $W(k,n), Y(k,n), Z(k,n), X(k,n)$, where $k$ and $n$ represents the frequency and the time indices, respectively.
An energy analysis of the components allows to estimate the diffuseness and DOA of the sound field in each TF-tile.
The DirAC parameters along with the omnidirectional component of FOA, $W$, represents the spatial audio scene. 

In the DirAC synthesis stage, the synthesized sound field originates from two streams, the directional and the diffuse streams. 
The directional stream is produced by generating point sources using amplitude panning of $W$, which is done in the original formulation of DirAC using Vector Base Amplitude Panning (VBAP)~\cite{Pulkki_vbap_1997}. 
The diffuse stream is responsible for the sensation of envelopment and is produced by conveying to the loudspeakers mutually decorrelated signals derived from $W$.
The original formulation of DirAC is aimed at synthesizing loudspeaker or binaural\cite{Laitinen_2009} outputs and is not directly applicable to Ambisonic coding.

\section{Methods}

\subsection{System overview}

The overall Ambisonic coding scheme is illustrated in Fig.~\ref{fig:overview}. 
At the encoder, the Ambisonic signal is conveyed along two paths.

In one branch, the DirAC parameters, namely the directions of arrival (DOAs) and the diffuseness values, are estimated after applying a Complex-valued Low-Delay Filterbank (CLDFB)~\cite{Schnell2008MPEG4EL} to the FOA extracted from the input signal.
CLDFB as employed produces complex-valued coefficients with a frequency resolution of \SI{400}{Hz} and a granularity of \SI{1.25}{ms}.
It has been found that this compromise in frequency resolution is still acceptable, even at low frequencies, and high temporal resolution is beneficial for synthesis.
The DirAC analysis is followed by a parameter coder, which quantizes and encodes the DirAC parameters to obtain a low bitrate parametric representation.
Since the DirAC analysis is performed in parallel to the core-coder, it has no impact on the encoder delay.

In the other branch, a truncated and passive downmixing or static matrixing is used to select a limited number of transport channels (TCs) to code.
Depending on the bitrate, 1 to 4 TCs are coded by the core-coder as specified in Tab.~\ref{tab:config}. 
At lower bitrates, the mono compatible omnidirectional component $W$ of HOA is coded by a mono coder. 
In this work the mono coder of 3GPP IVAS~\cite{3GPP253} is employed, which is a flexible version of the 3GPP EVS~\cite{3GPP445} accepting arbitrary bitrates.
In the case of 2 TCs, the left and right cardioid beam patterns are obtained from the $W$ and $Y$ components of HOA. 
It makes efficient use of the stereo coding mode of IVAS.
Finally, at higher bitrates, the FOA is extracted and transformed to A-format for being able to build channel pairs in the spatial domain for the stereo coding of IVAS.
The bitrate allocated to the core-coder is deducted from the overall target bitrate minus the bitrate needed for the DirAC parameters, achieving a Constant Bit-Rate (CBR) across frames.
The coded DirAC parameters and the coded audio bitstream are multiplexed before being transmitted over the channel.

At the decoder side, the decoded TCs are fed to a CLDFB analysis to perform the DirAC synthesis.
In the case of 4 TCs, and at higher bitrates, the low-frequency FOA can be recovered faithfully enough to perform the DirAC analysis on the decoder side on the decoded channels, saving the transmission of DirAC parameters.
However, at high frequencies, and due to the use of parametric bandwidth extension in the core-coder, the DirAC parameters are always transmitted in the high bands, i.e. above \SI{8}{kHz}.
The HOA output is synthesized using three sources: The two conventional DirAC streams, the directional and diffuse streams, and part or all of the first-order Ambisonic channels, which can be extracted directly from the TCs.
More specifically, for 1 TC the omni $W$ is directly ouputted, for 2 TCs, $W$ and $Y$, and for 4 TCs, the whole FOA. An energy compensation gain $g$ is applied to the directly extracted HOA channels for compensating the limitation of the two other streams as it will be explained later.
The obtained HOA signal is finally transformed back to the time domain.
Optionally, and depending on the application, other Ambisonic processing steps can be performed in TF domain, such as HOA decoding or binaural rendering, which can benefit from fast convolution algorithms in the frequency domain.

CLDFB analysis and synthesis on the decoder side add \SI{5}{ms} to the core-coder delay, for a total of \SI{37}{ms}.

\subsection{Parameter estimation}

The DirAC parameters are computed for non-overlapping and non-uniform frequency bands following roughly a multiple of the Equivalent Rectangular Bandwidth (ERB)~\cite{Glasberg_1990}.
A scale of about 8 times ERB is used to obtain from 5--6 parameter bands depending on the bitrate (cf. Tab.~\ref{tab:config}).

In each parameter band $b$ and time slot $n$, the DOA and the diffuseness of the sound field are estimated by computing a global intensity vector from the pressure $P$ and a velocity vector $\vec{U}$:
\begin{equation}
    \vec{\vect{I}}(b,n) = \sum_{k\in K_b} \text{Re} (P(k,n)\vec{U}^*(k,n)),
\end{equation}
where $K_b$ are the frequency indices of the parameter band $b$, $^*$ the complex conjuguate operator, and where:
\begin{equation}
    \left\{
    \begin{array}{ll}
        P(k,n)=W(k,n)\\
        \vec{U}(k,n)=X(k,n) \vec{e}_x+Y(k,n) \vec{e}_y+Z(k,n) \vec{e}_z 
    \end{array}
    \right.,
\end{equation}
where $\vec{e}_x,\vec{e}_y,\vec{e}_z$ represent the Cartesian unit vectors.

The energy of the sound field and the direction vector in the band $b$ can be estimated as:
\begin{equation}
    \left\{
    \begin{array}{ll}
        E(b,n)=\frac{1}{2}(\|\vect{P}(b,n)\|^2 + \|\vec{\vect{U}}(b,n)\|^2)\\
        \vec{dv}(b,n) = -\frac{\vec{\vect{I}}(b,n)}{\|\vec{\vect{I}}(b,n)\|}
    \end{array}.
    \right.
\end{equation}
The directional vector $dv$ is defined in Cartesian coordinates but can be easily transformed in spherical coordinates defined by a unity radius, the azimuth angle $\phi$ and elevation angle $\theta$.
The diffuseness is defined as the 1 minus the ratio between the expectations of the intensity vector L2-norm and the power of the sound field. 
It is estimated in our implementation by a moving average of size $P$ time slots, corresponding to a window size of \SI{32}{ms}:
\begin{equation}
    \Psi(b,n) = 1-\frac{\left \rVert \sum_{p=0}^{P} I(b,n-p) \right \rVert}{\sum_{p=0}^{P} E(b,n-p)}.
\end{equation}


\subsection{DirAC parameters coding}

Even if the number of DirAC parameters is calculated in a limited number of parameter bands, this still generates too much data for low bitrates.
The number of parameters is further reduced by downsampling them across time. 
The downsampling is decoupled between the diffuseness values and DOAs, exploiting the fact that diffuseness retains a longer term characteristic of the sound field than the DOA, which is a more reactive spatial cue.
The diffuseness is averaged over \SI{20}{ms}, while the DOAs are sent every \SI{5}{ms}.

The diffuseness is limited between 0 and 1 and non-uniformly quantized over \SI{3}{bits}. 
The thresholds and quantizer levels were derived from the well-known properties of the quantization sensitivity of inter-channel coherence (ICC) in multichannel coding, which is more sensitive to quantization error for coherence of 1 than for coherence of 0~\cite{Breebaart_2005}.
Diffuseness, which can be considered complementary in nature, is quantized finer for low values.
The 3-bit quantized indices are coded individually, but the bitrate can be reduced if the values across frequency bands are equal or very similar.

\begin{figure}[ht]
    \centering
    \includegraphics[width=\linewidth]{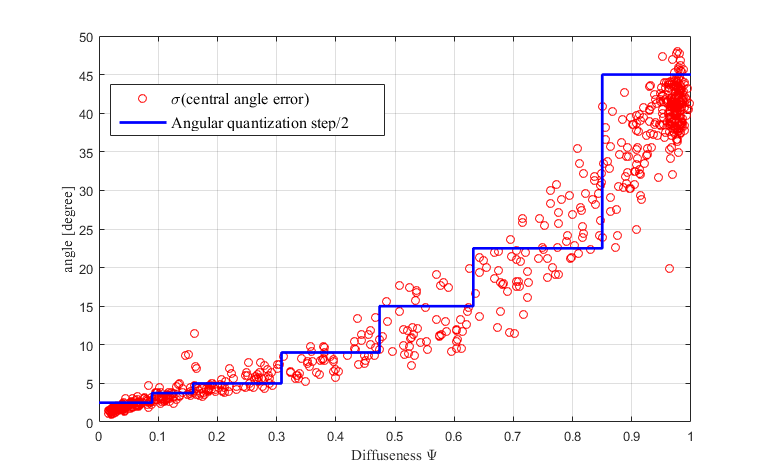}
    \caption{Standard deviation of the central angle between estimated DOAs and actual values in function of the diffuseness. The derived half quantization step function of the diffuseness is also indicated.}
    \label{fig:angularQ}
\end{figure}

Azimuth and elevation angles are quantized jointly and uniformly over the entire sphere, without any prior assumptions about DOA. 
Fig.~\ref{fig:angularQ} shows how the quantization resolution is derived from the standard deviation of the measured central angle between the estimated DOA and its actual value, as a function of diffuseness.
It can be observed that the estimated DOAs become inaccurate as the diffuseness increases, which corresponds to the perception that in a diffuse sound field, DOAs are no longer relevant.
As a result, the azimuth and elevation angles are quantized jointly between 2 (very diffuse) and 11 (non-diffuse) bits. 
Quantization is performed by spherical indexing as in~\cite{ragot_2023}, and additional delta coding and entropy coding could be used where advantageous over raw coding.

As indicated in Tab.~\ref{tab:config} and to avoid excessive side-information from the DirAC parameters, bit demand can be hard limited to a maximum by progressively lowering the resolution of the quantization.

\begin{table}[htbp]
    \caption{Main DirAC-based system configurations.}
    \centering
    \begin{tabular}{lcccc}
    \toprule
    Bit-rates    & \# TCs  &\# parameter bands & Max. DirAC bitrates   \\
    \midrule
    13.2--32 kbps & 1    & 5  & 3.5--\SI{8.5}{kbps}            \\ 
    48--96 kbps  & 2     & 5   &  6--\SI{10}{kbps}                \\
    96--128 kbps & 4     & 6 &  \SI{4.8}{kbps} (2 high bands)            \\ \bottomrule
    \end{tabular}
    \label{tab:config}
\end{table}

\subsection{Synthesis}

The HOA synthesis is divided into three groups. 
The first Ambisonic channels up to the order $L_{1}$ and degree $M_{1}$ can be directly derived from the decoded TCs.
The higher order and degree channels are synthesized from both the directional and diffuse streams up to the truncation order $L$.
Beyond that, for even higher orders up to $H$, only the directional component is modeled.
Omitting the frequency and time indices for clarity, the synthesis of the HOA output $\tilde{B}_l^m$ can be summarized as:
\begin{equation}
    \tilde{B}_l^m=
    \left\{
    \begin{array}{ll}
        g.\hat{B}_l^m   & \text{if } l,m \leq L_{1}, M_{1}\\
        g.g_{dir}\hat{B}_0^0 + g.g_{diff}D(\hat{B}_0^0)  & \text{else if } l \leq L\\
        g_{dir}\hat{B}_0^0 & \text{otherwise}
    \end{array}
    \right.
\end{equation}
where $\hat{ }$ denotes the coding operator, $D$ a decorrelation, $g$ an energy compensation gain justified later, and $g_{dir}$ and $g_{diff}$ are the directional and diffuse gains, respectively, and are given by:
\begin{equation}
    \left\{
    \begin{array}{ll}
        g_{dir} = \sqrt{1-\hat{\Psi}}Y_l^m(\phi, \theta)\\
        g_{diff} = \sqrt{\frac{\hat{\Psi}}{2l+1}}
    \end{array}
    \right.
\end{equation}
The directional gain is deduced by evaluating the spherical harmonic values at the decoded DOA, while the denumerator of $g_{diff}$ is explained by the SN3D normalization.

It was found empirically that for typical loudspeaker layouts and for binauralization, modeling the diffuse field up to \SI{8}{kHz} and $L=1$ was sufficient for 1 and 2 TCs, while for 4 TCs it is extended up to $L=2$.
The frequency limitation can be explained by the fact that phase is perceptually less relevant at high frequencies, where decorrelation plays a minor role. 
The order limitation is justified by the dominance of the spatial resolution of point sources in the perception of spatial audio quality.
Since the diffuse component is only modeled until the order $L$, an energy loss of the synthesized sound field is expected. 
It is counterbalanced by the energy preservation gain $g$.
The gain is derived from the observation that, following the DirAC model, the overall energy of the synthesized sound field by limiting to order $L$ the modeling of diffuse component decreases from: 
\begin{equation}
    \label{eq:nrg1}    \sum_{l=0}^H\sum_{m=-l}^l \|\bar{B}_l^m\|^2 = \|\hat{B}_0^0\|^2(H+1)
\end{equation}
to
\begin{equation}
    \label{eq:nrg2}
    \sum_{l=0}^H\sum_{m=-l}^l \|\tilde{B}_l^m\|^2 = \|\hat{B}_0^0\|^2 \left( g^2(L+1) + (H-L)(1-\Psi) \right).
\end{equation}
Making (\ref{eq:nrg2}) equal to (\ref{eq:nrg1}) gives then:
\begin{equation}
    g = \sqrt{1-\Psi\left( \frac{H+1}{L+1}-1 \right)}.
\end{equation}

\section{Evaluation}
\begin{figure}[t]
    \centering
    \includegraphics[width=\linewidth]{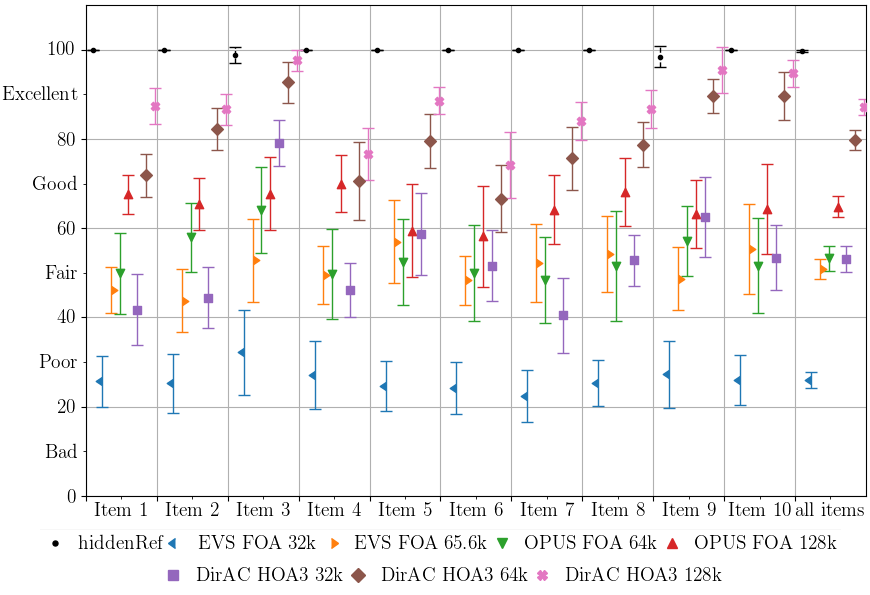}
    \caption{MUSHRA listening test results on HOA3 and FOA coding from 32 to \SI{128}{kbps}, with 95\% confidence intervals.}
    \label{fig:mushra}
\end{figure}

\subsection{Complexity}

The algorithmic complexity is evaluated in an objective manner and measured in weighted million operations per second (WMOPS). 
Complexity is reported in Tab.~\ref{tab:complexity} for multi-mono EVS coding individual FOA channels in comparison with the total complexity of the DirAC-based Ambisonic coder.
The DirAC-based system is significantly less complex than the multi-mono approach, even when synthesizing 16 output channels for the 3\textsuperscript{rd} order HOA (HOA3).

\begin{table}[htbp]
    \caption{Worst case complexity measured in WMOPS at \SI{32}{kHz} I/O.}
    \centering
    \begin{tabular}{lrrr} \toprule
    Conditions   & Encoder & Decoder FOA & Decoder HOA3  \\\midrule
    4x EVS at \SI{16.4}{kbps}         & 246.2    & 114.2 & -- \\
    DirAC 1-TC  \SI{32}{kbps}                & 81.3  & 69.2 & 118.1\\
    DirAC 2-TCs \SI{64}{kbps}                 & 85.2  & 69.3 & 119.2\\
    DirAC 4-TCs \SI{128}{kbps}               & 140.8  & 60.7 & 171.2\\ \bottomrule
    \end{tabular}
    \label{tab:complexity}
\end{table}

\subsection{Quality: Experimental setup}

To assess the subjective quality, a MUSHRA listening test compliant with ITU-R BS.1534~\cite{ITU-R_BS1534-3} was carried out. 
Thirteen expert listeners participated in the test using professional headphones and soundcards.
For evaluating diverse scenarios, the test items described in Tab.~\ref{tab:items} were considered.
The FOA and HOA3 signals were binauralized for static headphones playback using the Magnitude Least-Squares (MagLS) binaural rendering~\cite{Schoerkhuber_2018} and the Sadie II KU100 HRTFs~\cite{sadie2_2018}.
The proposed system tested at 32, 64 and \SI{128}{kbps} was compared to the reference codecs, EVS~\cite{3GPP445} and OPUS~\cite{rfc6716}, operating in multi-mono and CMF 3 mode, respectively.
Multi-mono EVS was used up to 4 times \SI{16.4}{kbps}, since individual channel coding is clearly less efficient at higher bitrates than the coupling performed in CMF 3 and the exploitation of OPUS stereo.
OPUS CMF 3 was only tested from \SI{64}{kbps} upwards, since below OPUS stereo tends to code a mono downmix signal only, and thus destroys the Ambisonics channel relationship.

\begin{table}[htbp]
    \caption{Test item description.}
    \centering
    \begin{tabular}{lll} 
    \toprule
    Indices & Recording type & Description \\
    \midrule
    3,9,10          & Synthetic   & 1--2 speakers with and w/o reverberation \\
    1          & EigenMike      & Outdoor conversation with traffic \\
    2          & EigenMike      & Indoor conversation in cafeteria \\
    4--5       & EigenMike      & 2--3 speakers \\
    6          & EigenMike      & 1 moving speaker  \\
    7          & EigenMike      & 2 speakers on the street  \\
    8          & EigenMike      & 4 speakers in office  \\
    \bottomrule
    \end{tabular}
    
    \label{tab:items}
\end{table}

\subsection{Quality: Results and discussion}

As shown in Fig.~\ref{fig:mushra} DirAC-based coder scales very well with the bitrate and outperforms clearly the other conditions under test at the same bitrates (by over 20 MUSHRA points).
It is particularly true for the 3 first synthetic test items, generated by Ambisonic panning of audio objects and for item 3 simulating reverberation, where even the 1 TC condition delivers good to very good quality.
However, for more realistic and complex audio scenes, which could violate the assumption of a single directional source in one parameter band, DirAC 1 TC creates more audible artifacts, such as an unstable spatial image or unnatural ambience.
Nonetheless, it saves around \SI{50}{\%} of the bitrate compared to multi-mono EVS at \SI{65.4}{kbps}.
The transmission of 2 TCs, or even more 4 TCs, reduces the contribution of decorrelators needed, resulting in a much more faithful reproduction of the diffuse sound. 
It also stabilizes the synthesized point sources.

\section{Conclusion}
\label{sec:conclude}
In this paper, a first-order DirAC-based Ambisonic coder was proposed in the realm of real-time speech communications. 
It has been demonstrated that, through the choices and techniques proposed, an efficient coding of HOA can be achieved, even under the constraints of low bitrates and low delay.
Test results on diverse immersive communication scenarios highlight the merit of such a parametric coding for HOA.
However, the quality achieved by a first-order DirAC model saturates at higher bitrates, above \SI{128}{kbps}, where higher orders of the paradigm (HO-DirAC) or waveform-preserving schemes, such as prediction and joint channel coding, could claim to scale towards perceptual transparency.
A combination of these latter approaches and the proposed DirAC-based solution has been adopted in the recently standardized 3GPP IVAS codec. 

\section*{Acknowledgement}
The authors would like to thank J{\"u}rgen Herre, Stefan Bayer, Srikanth Korse and Markus Multrus for their insightful input and guidance.

\vfill
\pagebreak
\IEEEtriggeratref{14}
\bibliographystyle{IEEEtran}
\bibliography{IEEEabrv,refs}

\begin{thebibliography}{10}
\providecommand{\url}[1]{#1}
\csname url@samestyle\endcsname
\providecommand{\newblock}{\relax}
\providecommand{\bibinfo}[2]{#2}
\providecommand{\BIBentrySTDinterwordspacing}{\spaceskip=0pt\relax}
\providecommand{\BIBentryALTinterwordstretchfactor}{4}
\providecommand{\BIBentryALTinterwordspacing}{\spaceskip=\fontdimen2\font plus
\BIBentryALTinterwordstretchfactor\fontdimen3\font minus
  \fontdimen4\font\relax}
\providecommand{\BIBforeignlanguage}[2]{{%
\expandafter\ifx\csname l@#1\endcsname\relax
\typeout{** WARNING: IEEEtran.bst: No hyphenation pattern has been}%
\typeout{** loaded for the language `#1'. Using the pattern for}%
\typeout{** the default language instead.}%
\else
\language=\csname l@#1\endcsname
\fi
#2}}
\providecommand{\BIBdecl}{\relax}
\BIBdecl

\bibitem{opus_ambix}
\BIBentryALTinterwordspacing
J.~Skoglund and M.~Graczyk, ``{Ambisonics in an Ogg Opus Container},'' {…RFC}
  8486, Oct. 2018. [Online]. Available:
  \url{https://tools.ietf.org/html/rfc8486.html}
\BIBentrySTDinterwordspacing

\bibitem{McGrath_2019}
D.~McGrath, S.~Bruhn, H.~Purnhagen, M.~Eckert, J.~Torres, S.~Brown, and
  D.~Darcy, ``{Immersive Audio Coding for Virtual Reality Using a
  Metadata-assisted Extension of the 3GPP EVS Codec},'' in \emph{{ICASSP 2019 -
  2019 IEEE International Conference on Acoustics, Speech and Signal Processing
  (ICASSP)}}, 2019, pp. 730--734.

\bibitem{peters2016scene}
N.~Peters, D.~Sen, M.-Y. Kim, O.~Wuebbolt, and M.~S.~Weiss, ``Scene-based audio
  implemented with higher order ambisonics,'' \emph{SMPTE Motion Imaging
  Journal}, vol. 125, no.~9, pp. 16--24, 2016.

\bibitem{Zamani_2017}
S.~Zamani, T.~Nanjundaswamy, and K.~Rose, ``Frequency domain singular value
  decomposition for efficient spatial audio coding,'' in \emph{{2017 IEEE
  Workshop on Applications of Signal Processing to Audio and Acoustics
  (WASPAA)}}, 2017, pp. 126--130.

\bibitem{mahe_2019}
P.~Mah{\'e}, S.~Ragot, and S.~Marchand, ``{First-order ambisonic coding with
  PCA matrixing and quaternion-based interpolation},'' in \emph{{Conference:
  22nd International Conference on Digital Audio Effects (DAFx19)}}, Sep. 2019.

\bibitem{xu2021higher}
x.~Jiahao, n.~yadong, w.~xihong, and q.~tianshu, ``higher order ambisonics
  compression method based on independent component analysis,'' \emph{{Journal
  of the Audio Engineering Society}}, no. 10456, May 2021.

\bibitem{Bleidt_2017}
R.~L. Bleidt, D.~Sen, A.~Niedermeier, B.~Czelhan, S.~F{\"u}g, S.~Disch,
  J.~Herre, J.~Hilpert, M.~Neuendorf, H.~Fuchs, J.~Issing, A.~Murtaza,
  A.~Kuntz, M.~Kratschmer, F.~K{\"u}ch, R.~F{\"u}g, B.~Schubert, S.~Dick,
  G.~Fuchs, F.~Schuh, E.~Burdiel, N.~Peters, and M.-Y. Kim, ``{Development of
  the MPEG-H TV Audio System for ATSC 3.0},'' \emph{IEEE Transactions on
  Broadcasting}, vol.~63, no.~1, pp. 202--236, 2017.

\bibitem{pulkki2007}
V.~Pulkki, ``{spatial sound reproduction with directional audio coding},''
  \emph{{Journal of the Audio Engineering Society}}, vol.~55, pp. 503--516,
  Jun. 2007.

\bibitem{Politis_2015}
A.~Politis, J.~Vilkamo, and V.~Pulkki, ``{Sector-Based Parametric Sound Field
  Reproduction in the Spherical Harmonic Domain},'' \emph{{IEEE Journal of
  Selected Topics in Signal Processing}}, vol.~9, no.~5, pp. 852--866, 2015.

\bibitem{Hold_2024}
C.~Hold, V.~Pulkki, A.~Politis, and L.~McCormack, ``{Compression of
  Higher-Order Ambisonic Signals Using Directional Audio Coding},''
  \emph{{IEEE/ACM Transactions on Audio, Speech, and Language Processing}},
  vol.~32, pp. 651--665, 2024.

\bibitem{Hold_icassp_2024}
C.~Hold, L.~McCormack, A.~Politis, and V.~Pulkki, ``{Perceptually-Motivated
  Spatial Audio Codec for Higher-Order Ambisonics Compression},'' in
  \emph{{ICASSP 2024 - 2024 IEEE International Conference on Acoustics, Speech
  and Signal Processing (ICASSP)}}, 2024, pp. 1121--1125.

\bibitem{hirvonen2009perceptual}
T.~Hirvonen, J.~Ahonen, and V.~Pulkki, ``Perceptual compression methods for
  metadata in directional audio coding applied to audiovisual teleconference,''
  \emph{{Journal of the Audio Engineering Society}}, no. 7706, May 2009.

\bibitem{3GPP445}
\BIBentryALTinterwordspacing
``{Codec for Enhanced Voice Services (EVS); Detailed algorithmic
  description},'' 3GPP, TS 26.445, May 2024. [Online]. Available:
  \url{https://www.3gpp.org/DynaReport/26445.htm}
\BIBentrySTDinterwordspacing

\bibitem{3GPP253}
\BIBentryALTinterwordspacing
``{Codec for Immersive Voice and Audio Services (IVAS); Detailed Algorithmic
  Description including RTP payload format and SDP parameter definitions},''
  3GPP, TS 26.253, Jul. 2024. [Online]. Available:
  \url{https://www.3gpp.org/DynaReport/26253.htm}
\BIBentrySTDinterwordspacing

\bibitem{rfc6716}
\BIBentryALTinterwordspacing
J.-M. Valin, K.~Vos, and T.~B. Terriberry, ``{Definition of the Opus Audio
  Codec},'' {RFC} 6716, Sep. 2012. [Online]. Available:
  \url{https://www.rfc-editor.org/info/rfc6716}
\BIBentrySTDinterwordspacing

\bibitem{zotter2019ambisonics}
\BIBentryALTinterwordspacing
F.~Zotter and M.~Frank, \emph{{Ambisonics: A Practical 3D Audio Theory for
  Recording, Studio Production, Sound Reinforcement, and Virtual Reality}},
  ser. Springer Topics in Signal Processing.\hskip 1em plus 0.5em minus
  0.4em\relax Springer International Publishing, 2019. [Online]. Available:
  \url{https://books.google.de/books?id=v7rewgEACAAJ}
\BIBentrySTDinterwordspacing

\bibitem{Nachbar2011AMBIXA}
\BIBentryALTinterwordspacing
C.~Nachbar, F.~Zotter, E.~Deleflie, and A.~Sontacchi, ``{AMBIX - A Suggested
  Ambisonics Format},'' 2011. [Online]. Available:
  \url{https://api.semanticscholar.org/CorpusID:4238976}
\BIBentrySTDinterwordspacing

\bibitem{Pulkki_vbap_1997}
V.~Pulkki, ``\BIBforeignlanguage{English}{{Virtual Sound Source Positioning
  Using Vector Base Amplitude Panning}},''
  \emph{\BIBforeignlanguage{English}{{Journal of the Audio Engineering
  Society}}}, vol.~45, no.~6, pp. 456--466, 1997.

\bibitem{Laitinen_2009}
M.-V. Laitinen and V.~Pulkki, ``{Binaural reproduction for Directional Audio
  Coding},'' in \emph{{2009 IEEE Workshop on Applications of Signal Processing
  to Audio and Acoustics}}, 2009, pp. 337--340.

\bibitem{Schnell2008MPEG4EL}
\BIBentryALTinterwordspacing
M.~Schnell, M.~Schmidt, M.~Jander, T.~Albert, R.~Geiger, V.~T. Ruoppila,
  P.~Ekstrand, and B.~Grill, ``{MPEG-4 Enhanced Low Delay AAC - A New Standard
  for High Quality Communication},'' \emph{{Journal of The Audio Engineering
  Society}}, 2008. [Online]. Available:
  \url{https://api.semanticscholar.org/CorpusID:29023986}
\BIBentrySTDinterwordspacing

\bibitem{Glasberg_1990}
\BIBentryALTinterwordspacing
B.~R. Glasberg and B.~C. Moore, ``Derivation of auditory filter shapes from
  notched-noise data,'' \emph{Hearing Research}, vol.~47, no.~1, pp. 103--138,
  1990. [Online]. Available:
  \url{https://www.sciencedirect.com/science/article/pii/037859559090170T}
\BIBentrySTDinterwordspacing

\bibitem{Breebaart_2005}
J.~Breebaart, S.~van~de Par, A.~Kohlrausch, and E.~Schuijers, ``Parametric
  coding of stereo audio,'' \emph{{EURASIP Journal on Advances in Signal
  Processing}}, vol. 2005, pp. 1--18, 2005.

\bibitem{ragot_2023}
S.~Ragot and A.~Vasilache, ``{Spherical Vector Quantization for Spatial
  Direction Coding},'' in \emph{{ICASSP 2023 - 2023 IEEE International
  Conference on Acoustics, Speech and Signal Processing (ICASSP)}}, 2023, pp.
  1--5.

\bibitem{ITU-R_BS1534-3}
\emph{Method for the subjective assessment of intermediate quality level of
  audio systems}, {International Telecommunication Union, ITU} Std. BS.1534-3,
  2015.

\bibitem{Schoerkhuber_2018}
C.~Sch{\"o}rkhuber, M.~Zaunschirm, and R.~H{\"o}ldrich, ``{Binaural Rendering
  of Ambisonic Signals via Magnitude Least Squares},'' Mar. 2018.

\bibitem{sadie2_2018}
\BIBentryALTinterwordspacing
C.~Armstrong, L.~Thresh, D.~Murphy, and G.~Kearney, ``{A Perceptual Evaluation
  of Individual and Non-Individual HRTFs: A Case Study of the SADIE II
  Database},'' \emph{Applied Sciences}, vol.~8, no.~11, 2018. [Online].
  Available: \url{https://www.mdpi.com/2076-3417/8/11/2029}
\BIBentrySTDinterwordspacing

\end{thebibliography}

\end{document}